\documentclass{PoS}

\usepackage{amsfonts}
\usepackage{amsmath}
\usepackage{subfigure} % Include figure files

\title{
  Determination of $B_K$ using improved staggered fermions
  (IV) One-loop matching
}

\ShortTitle{
  Determination of $B_K$ using improved staggered fermions (IV)
}

\author{\speaker{Jangho Kim}, Taegil Bae, Hyung-Jin Kim,
  Jongjeong Kim, Kwangwoo Kim, Boram Yoon, Weonjong Lee\\
  Frontier Physics Research Division and Center for Theoretical Physics \\
  Department of Physics and Astronomy, 
  Seoul National University, Seoul, 151-747, South Korea \\
  E-mail: \email{wlee@snu.ac.kr}}

\author{Chulwoo Jung \\
  Physics Department, Brookhaven National Laboratory,
  Upton, NY11973, USA \\
  E-mail: \email{chulwoo@bnl.gov}}

\author{Stephen R. Sharpe\\
  Physics Department, University of Washington, Seattle, 
  WA 98195-1560, USA \\
  E-mail: \email{sharpe@phys.washington.edu}}

\abstract{ We discuss the impact of using one-loop matching
  on the calculation of $B_K$ using HYP-smeared improved staggered
 fermions. We give estimates of size of the truncation errors
 from the missing two-loop corrections.}

\FullConference{
  The XXVII International Symposium on Lattice Field Theory - LAT2009\\
  July 26-31 2009
  Peking University, Beijing, China
}

\begin{document}

\section{Introduction} 
This paper completes a series of four reports on our
calculation of $B_K$ using HYP-smeared staggered fermions.
In the previous reports we presented
the results of fitting using SU(3)~\cite{ref:wlee:2009-1}
and SU(2)~\cite{ref:wlee:2009-2} staggered
chiral perturbation theory, and our method for
estimating the systematic error due to 
finite volume effects~\cite{ref:wlee:2009-3}.
Here we focus on impact of the matching factors that
we use to connect the lattice operators 
to their continuum counterparts, 
and explain how we estimate the errors that are introduced by
truncating this matching at one-loop order.
Further details will be given in Ref.~\cite{ref:future}.

\section{One-loop Matching}
To define matching factors we need to specify the continuum 
regularization and renormalization scheme used to define the
operators in the continuum. If one matches perturbatively,
as we do here, it is conventional to use
$\overline{\rm MS}$ regularization
with the NDR (Naive Dimensional Regularization) prescription for $\gamma_5$.
One also needs to choose which class of lattice operators to use,
and we follow earlier work and
use the so-called two trace approach~\cite{oldBK}.
One then calculates the matrix elements of the $B_K$ operator
at some order (here one-loop)
in both continuum and lattice regularizations. Equating them
determines the matching factor---which is, in general, a matrix.

Alternatively one can use the RI-MOM scheme, which is defined
in any regularization, and determine the matching by a non-perturbative
calculation on the lattice. The advantages and disadvantages
of this scheme are reviewed in Ref.~\cite{ref:yasumichi}.
We ultimately plan to use it, but so far have only used this method for
bilinear operators~\cite{ref:andrew}.

Returning to the perturbative approach, the
one-loop matching factors $Z_{ij}$ are defined through
\begin{eqnarray}
  O_i^\text{Cont} (\mu) &=& Z_{ij} (\mu, a) O_j^\text{Latt}(a)
  \label{eq:match-1}
  \\
  Z_{ij} &=& \delta_{ij} + \frac{\alpha_s}{4\pi}
  [\gamma_{ij} \log(\mu a) + c_{ij}]
  \\
  c_{ij} &=& C^\text{Cont}_{ij} - C^\text{Latt}_{ij}
\end{eqnarray}
where $O_i^\text{Cont}(\mu)$ are continuum $\Delta S = 2$ operators
(here a single operator)
renormalized at scale $\mu$ and
$O_j^\text{Latt}(a)$ are the lattice operators required
for the matching.
$\gamma_{ij}$ is the anomalous dimension matrix,
while $C^\text{Cont}_{ij}$ and $C^\text{Latt}_{ij}$ are the finite
parts of the continuum and lattice matrix elements, respectively.
% 
%The $\alpha_s$ is the strong coupling constant ($\alpha_s=g^2/(4\pi)$).
%
The list of lattice operators which appear at one-loop order
is given in Ref.~\cite{ref:LSpert2}. This reference also calculates the
matching factors for the HYP-smeared operators we use, but with
the Wilson gauge action. The generalization to the Symanzik gauge
action used to generate the MILC configurations will be
presented in Ref.\cite{ref:wlee:2009-10}. 
The results we use here are preliminary.

In applying (\ref{eq:match-1}) we make one simplification: from the
rather long list of operators $O_j^\text{Latt}$ which contribute at
one-loop we keep only the four which have the same taste as the
external kaons ($\xi_5$). This introduces $O(\alpha/(4\pi))$
truncation errors which turn out to be of next-to-leading order (NLO)
in SU(3) staggered chiral perturbation theory
(SChPT)~\cite{ref:sharpe:1}, and of NNLO in SU(2)
SChPT~\cite{ref:wlee:2009-1,ref:wlee:2009-2,ref:future}.  Our SU(3)
fits attempt to pick out the contribution of the missing operators and
then remove them. Our SU(2) fits ignore these contributions as being
of too high order.

We now show how moving from tree-level to one-loop matching
impacts the results for $B_K$. Here we first use what we call
``parallel matching'', in which the scale in the continuum operator
is set to $\mu=1/a$, and in which $\alpha_s$ (which we take to
be in the $\overline{\rm MS}$ scheme) is also evaluated at this
scale. The rationale for this choice is that it is a reasonable
estimate for the typical momentum contributing in the matching.
It is also possible to estimate the scale to use
(usually called ``$q^*$'') based on the integrand of the
one-loop integrals, but we have not yet attempted this.

\begin{figure}[t!]
  \centering
  \includegraphics[width=0.49\textwidth]
                  {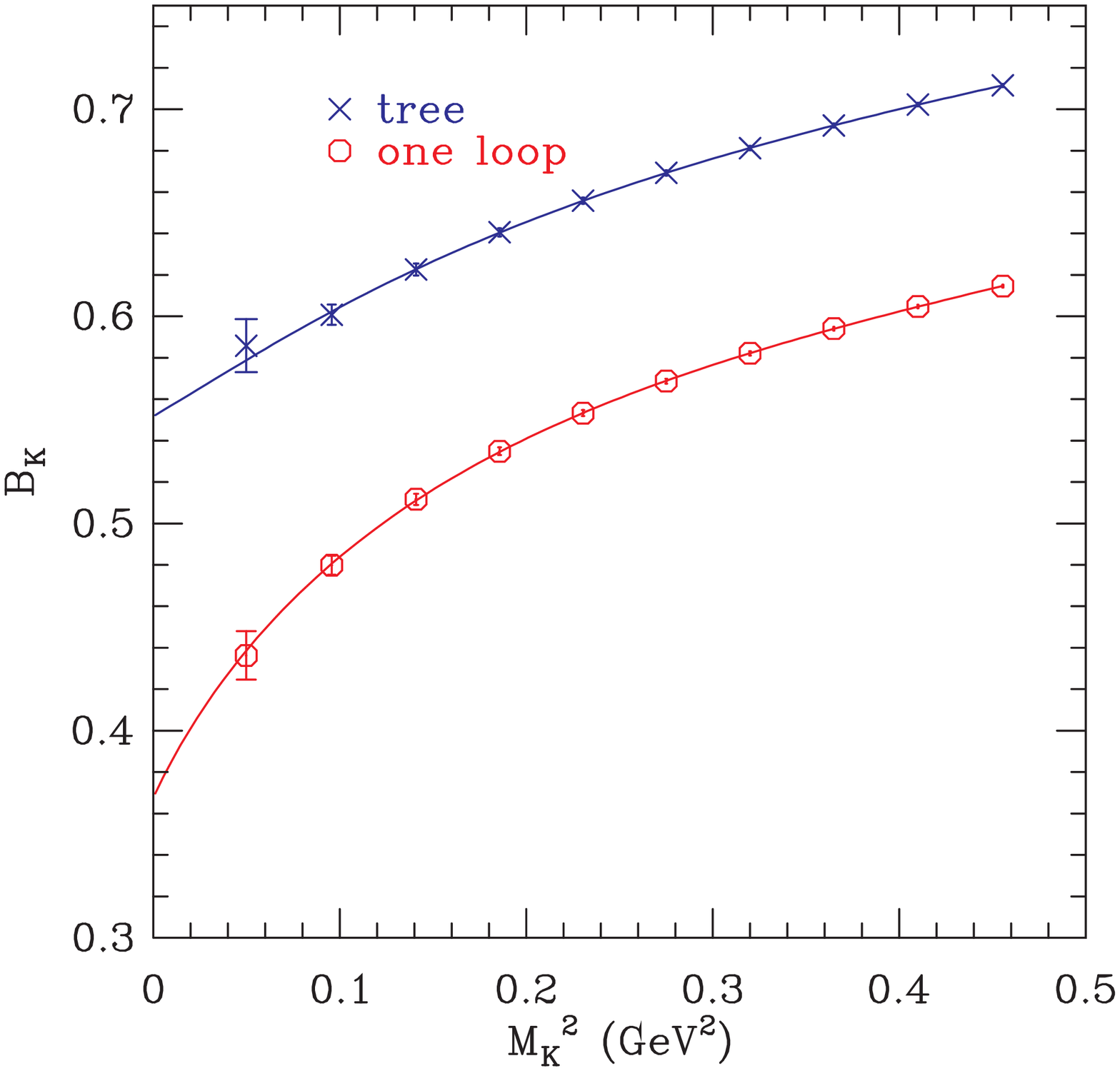}
  \includegraphics[width=0.49\textwidth]
                  {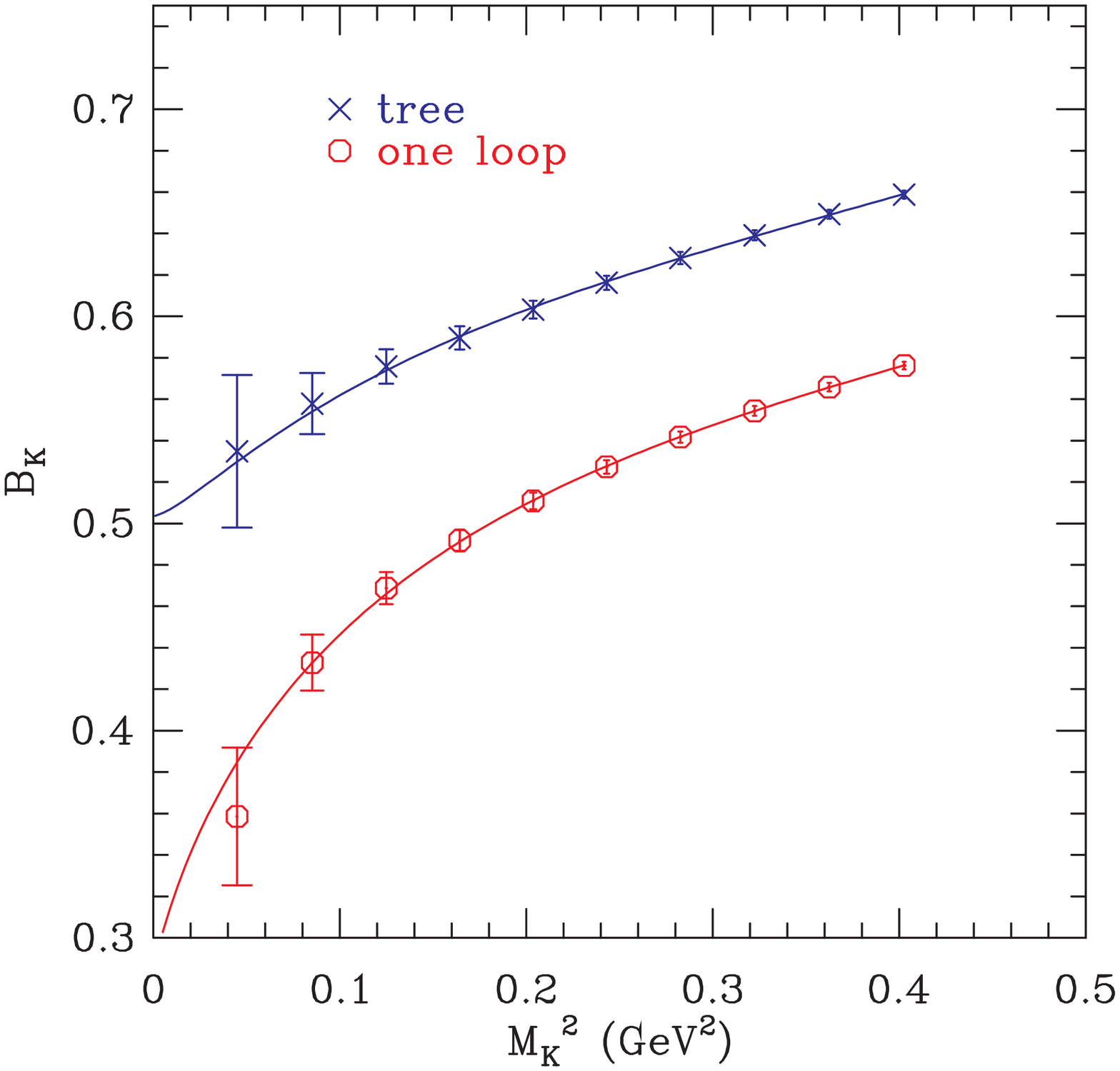}
\caption{Tree-level and one-loop matched $B_K$ versus $M_K^2$ 
for degenerate valence quarks. Results are for
the coarse (left) and fine (right) MILC lattices with 
$m_l/m_s = 0.01/0.05$.}
\label{fig:bk-C3-F1}
\end{figure}
In Fig.~\ref{fig:bk-C3-F1}, we show the results for $B_K$ on
coarse ($a\approx 0.12\;$fm) 
and fine ($a\approx 0.09\;$fm) 
lattices before
and after inclusion of the one-loop corrections. 
For clarity, we show only the points in which the valence quarks 
are degenerate. The roughly 20\% reduction caused by the
inclusion of one-loop contributions holds also for 
non-degenerate valence quarks.
We also show the results of a four-parameter partial NNLO fit.\footnote{%
Details of this fit which will be explained in Ref.~\cite{ref:future},
and are not pertinent here.}

The size of the one-loop shift ($\approx 20\%$) is of the 
expected magnitude,
given that $\alpha_s(1/a)$ is $\approx 0.33$ and $0.27$ on the
coarse and fine lattices, respectively.
It should be kept in mind that, however,
that $B_K$ is scale dependent, and so the one-loop correction can be
made larger or smaller by varying the scale chosen
in the continuum operator. In other words, there is no precise
way of defining the size of the correction.

\begin{figure}[t!]
  \centering
  \includegraphics[width=0.49\textwidth]
                  {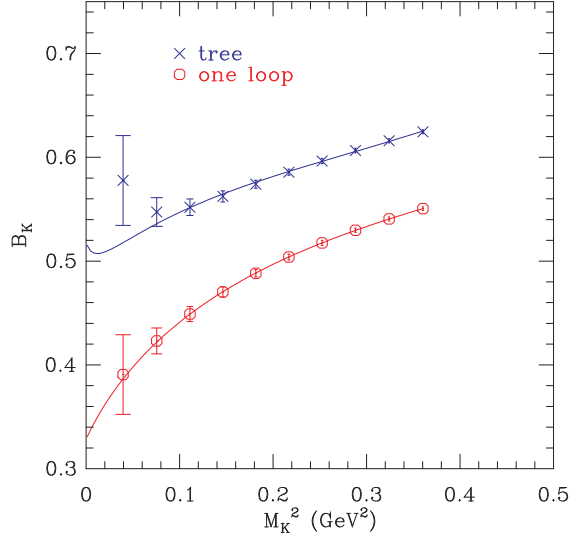}
\caption{ $B_K$ vs. $M_K^2$ for the MILC superfine lattice with
  $m_l/m_s = 1/5$.}
\label{fig:bk-S1}
\end{figure}

A noteworthy feature of these results is that the curvature at
small $M_K$ is larger after one-loop matching, This is even
more pronounced in the results on the superfine lattices ($a\approx 0.06\;$fm),
shown in Fig.~\ref{fig:bk-S1}.
Indeed, one can see that the fit to the tree-level results
has an upwards ``hook'' at very small $M_K$, which is the result
of the fit requiring a significant contribution from the taste-violating
operators which are present because of truncation (and discretization)
errors. These contributions behave as $-\log M_K$ in the chiral limit,
and are finite there because the kaon that appears has non-Goldstone
taste and so its mass does not vanish in the chiral limit.
We expect such contributions to be of leading order in SChPT for
tree-level matching, but of NLO, and thus much smaller, 
for one-loop matching. This is consistent
with our results. We stress that the curvature seen in
the one-loop curves is not a surprise as the chiral logarithm
has a fairly large coefficient. 

\begin{figure}[htbp]
  \centering
  \includegraphics[width=0.49\textwidth]
                  {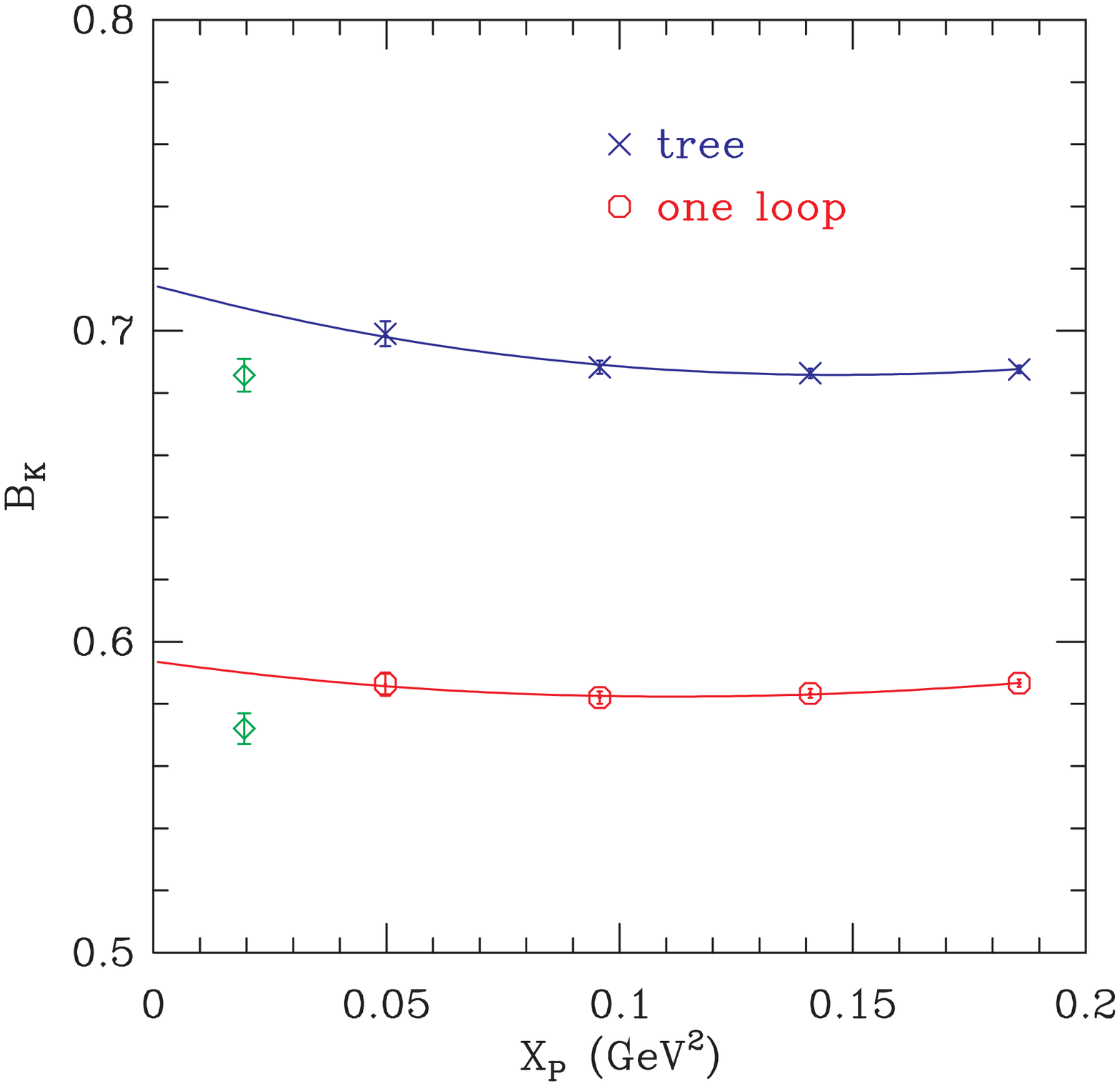}
  \includegraphics[width=0.49\textwidth]
                  {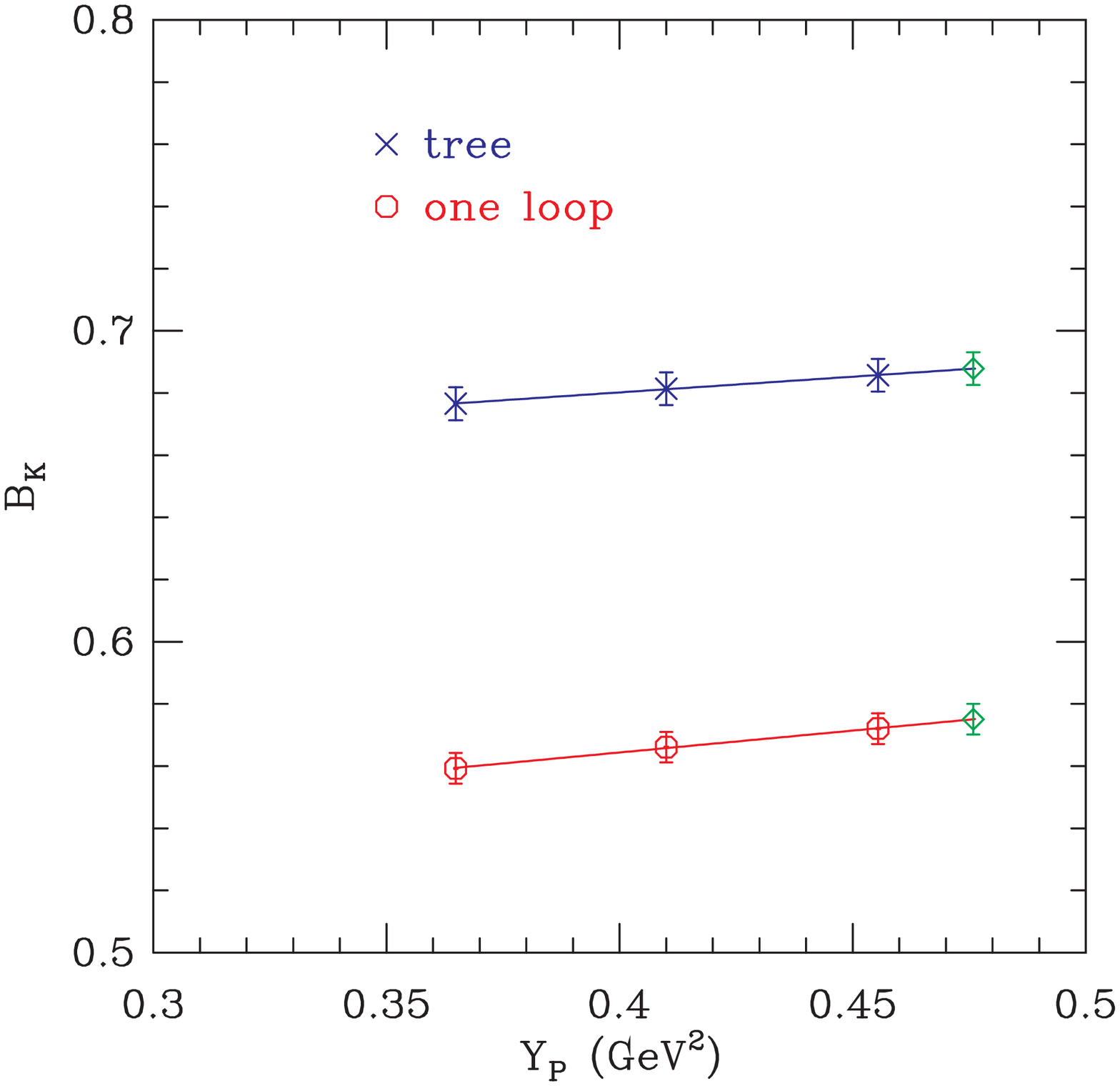}
\caption{ $B_K$ vs. $X_P$ (left: X-fit) and vs. $Y_P$ (right: Y-fit)
on the MILC coarse lattices (right: Y-fit)
with $m_l/m_s = 1/5$. The [green] diamonds in the
left plot show the result after extrapolation to the physical pion
mass, $X_P=M_\pi^2$, and removal of lattice artefacts from the fit
function. In the right panel the [green] diamonds show the result
of an extrapolation to the physical valence strange quark mass.}
\label{fig:bk-su2}
\end{figure}

The subset of the data most relevant to extrapolating to the physical
kaon is that in which the valence kaon is maximally non-degenerate.
This effect of one-loop matching on this subset of the data
is illustrated by Fig.~\ref{fig:bk-su2}, 
where we present the results of the SU(2) SChPT fits
$B_K$ on the coarse MILC lattices.
We show an example of the ``X-fit'' (the extrapolation in $M_\pi^2$
for fixed valence strange-quark mass) and
the ``Y-fit'' (the extrapolation in the valence strange quark mass).
These fits are explained in Ref.~\cite{ref:wlee:2009-2}.

\section{RG Evolution}
The fitting procedures described in
Refs.~\cite{ref:wlee:2009-1,ref:wlee:2009-2}
result in values for $B_K(1/a)$ on the three lattice spacings,
with taste-breaking discretization and truncation errors removed.
In order to compare these values we next run them 
from $1/a$ to a common scale, which we take to be $2\;$GeV.
Here, we use two-loop RG evolution
\begin{eqnarray}
B_K^\text{NDR}(p) &=&
\frac{ [ 1 - \frac{\alpha(q^*)}{4\pi } Z ] }
{ [ 1 - \frac{\alpha(p)}{4\pi} Z ] }
\bigg( \frac{\alpha(p)}{\alpha(q^*)} \bigg)^{ d^{(0)} }
B_K^\text{NDR}(q^*)
\nonumber \\
\label{eq:bk-rg-evol}
\\
Z &=& \frac{ \gamma^{(1)} }{2 \beta_0} -
d^{(0)} \frac{ \beta_1 }{ \beta_0 }
\nonumber \\
d^{(0)} &=& \frac{ \gamma^{(0)} }{ 2 \beta_0 }
\nonumber
\end{eqnarray}
where the anomalous dimension matrices 
$\gamma^{(i)}$, and the beta-function coefficients,
$\beta_i$, are given, e.g., in Ref.~\cite{buras-90-1}.

After RG running, we have values of $B_K({\rm NDR},\mu=2\ {\rm GeV})$ from the
three lattices spacings, which still contain discretization and truncation
errors. 
We attempt to remove the former by a linear extrapolation in $a^2$,
as described in Ref.~\cite{ref:wlee:2009-2}. 
As for the latter, we attempt to estimate these separately,
as we now describe.

\section{Estimate of Two-Loop Terms}
Let $B_K^{(i)}$ be the value of $B_K$ obtained
using parallel matching ($\mu=1/a$) at the $i$'th loop level,
and after extrapolation to the physical valence and sea-quark masses.
Then we can define $\Delta B_K^{(i)}$ as
\begin{equation}
\Delta B_K^{(i)} \equiv B_K^{(i-1)} - B_K^{(i)}\,.
\end{equation}
so that $\Delta B_K^{(i)}$ represents the shift due to
the $i$'th loop correction to $B_K$.
We know $B_K^{(0)}$ and $B_K^{(1)}$ and so we can calculate $\Delta
B_K^{(1)}$; the results are collected in Table~\ref{tab:dbk2-su3}
for the SU(3) fits, and Table~\ref{tab:dbk2-su2} for the SU(2) fits.
One estimate of $\Delta B_K^{(2)}$ is then
\begin{equation}
\Delta B_K^{(2)} \approx \Delta B_K^{(1)} \times \alpha_s(1/a)\,,
\end{equation}
with results also given in the Tables.

\begin{figure}[t!]
  \centering
  \includegraphics[width=0.49\textwidth]{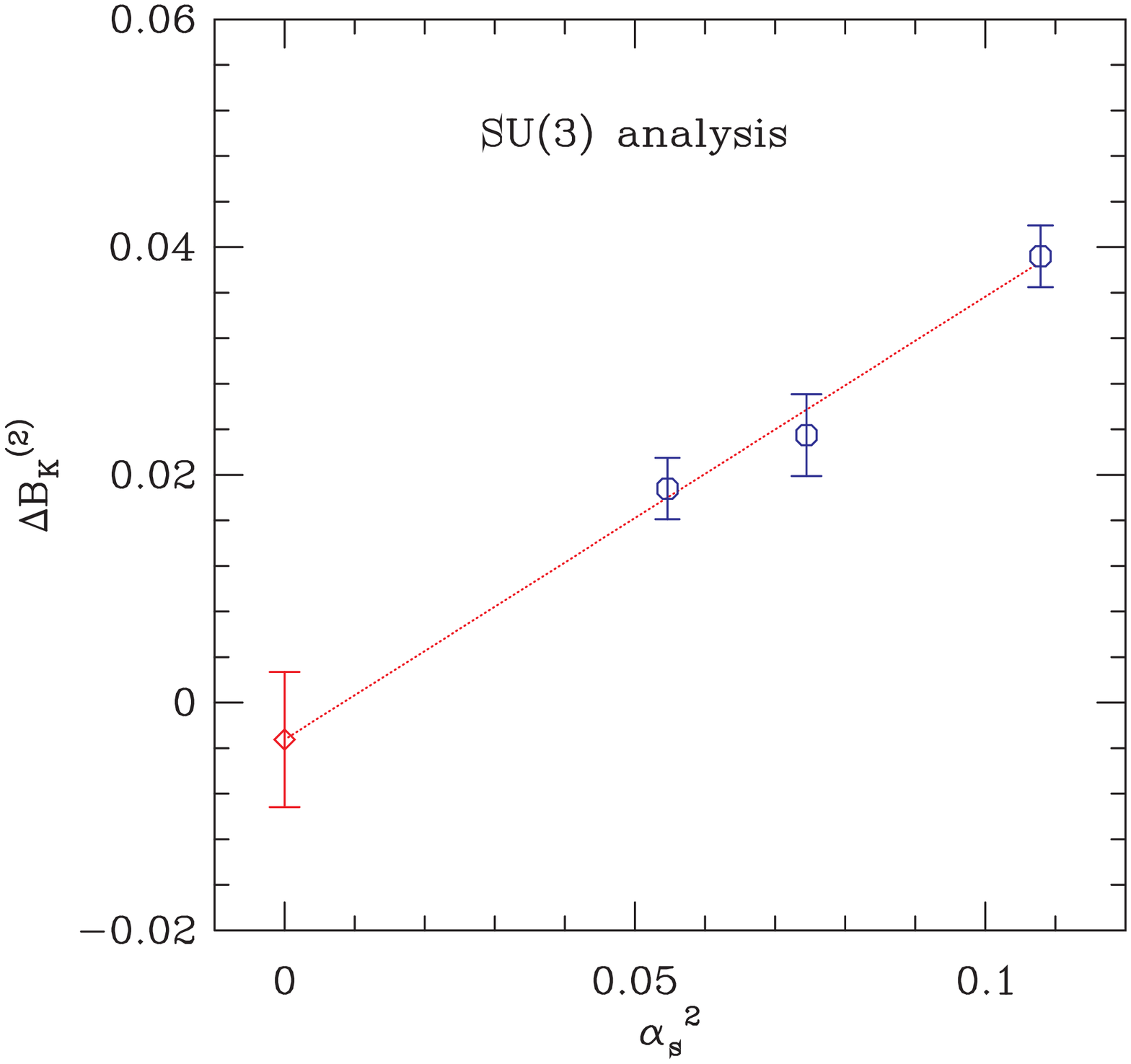}
  \includegraphics[width=0.49\textwidth]{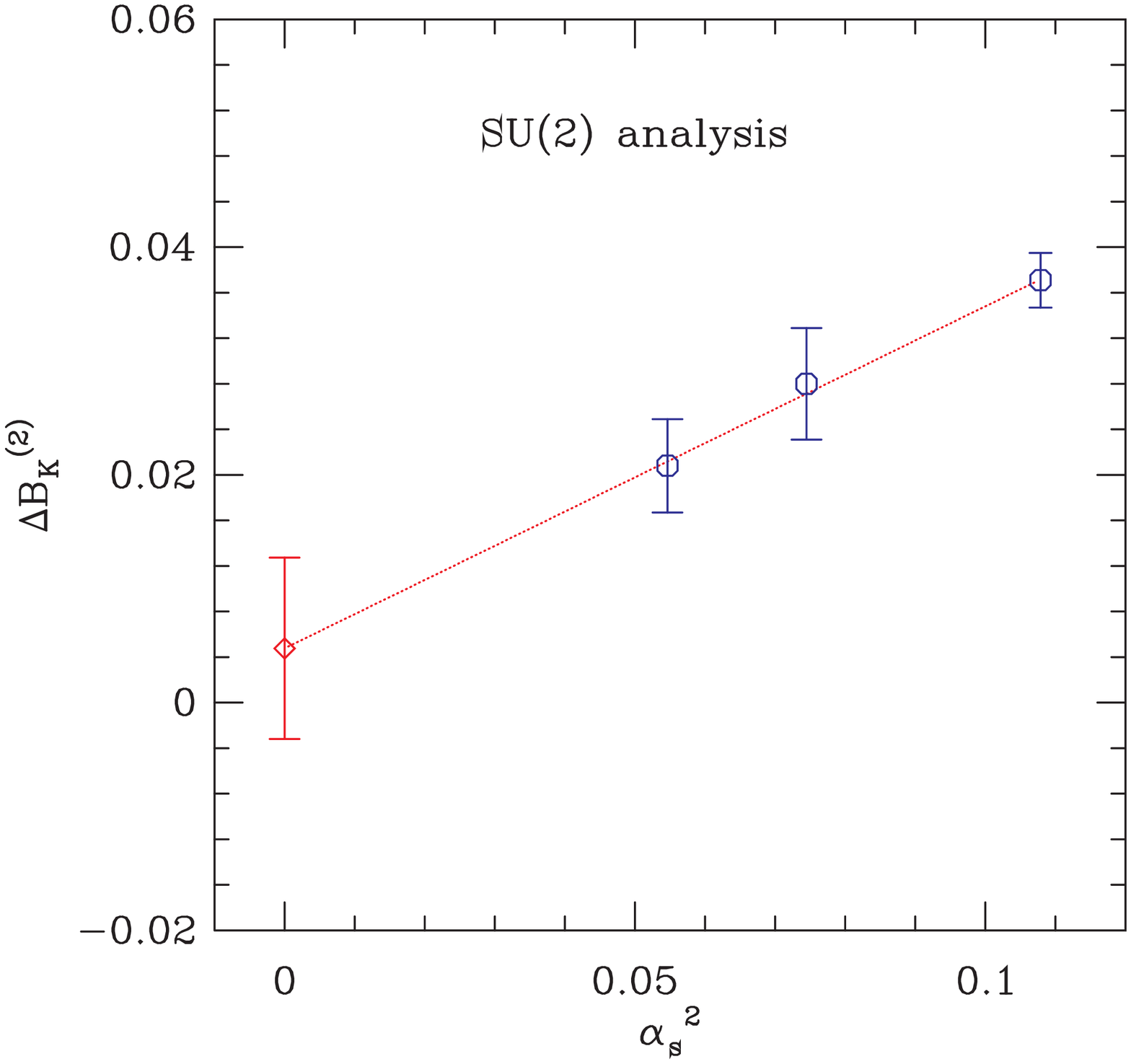}
\caption{ $\Delta B_K^{(2)}$ vs. $\alpha_s^2$ for the SU(3) analysis
  (left) and for the SU(2) analysis (right).}
\label{fig:dbk2-su3-su2}
\end{figure}

We plot $\Delta B_K^{(2)}$ versus $\alpha_s(1/a)^2$ 
for the two analyses in Fig.~\ref{fig:dbk2-su3-su2}.
Linear fits yield intercepts consistent with zero.
This is not surprising given that $\Delta B_K^{(1)}$ is
obtained from a one-loop matching formula, eq.~(\ref{eq:match-1}),
in which the correction is proportional to $\alpha_s(1/a)$,
and is then multiplied by  $\alpha_s(1/a)$ again to obtain
$\Delta B_K^{(2)}$. The vanishing of the intercept
is is not, however, an automatic result because
the one-loop matching is applied {\em before} fitting and extrapolating
the data, and involves contributions from lattice operators having
different dependence on the quark masses.
This means that $\Delta B_K^{(1)}$ need not 
be exactly linear in $\alpha_s(1/a)$.

An alternative estimate is simply to use 
\begin{equation}
\Delta B_K^{(2)'}=B_K^{(0)} \times \alpha_s(1/a)^2\,,
\end{equation}
i.e. the naive estimate of the two-loop contribution.
This turns out to be somewhat larger than $\Delta B_K^{(2)}$,
as shown in the Tables. It also varies more rapidly with
the lattice spacing.

Since we extrapolate to the continuum limit assuming
a linear dependence on $a^2$, the more slowly varying truncation
error does not extrapolate to zero, although it will be somewhat
reduced from the value at our smallest lattice spacing.
To be conservative however, we take the value of the
truncation error {\em for the superfine lattices},
and we use the larger of the two estimates, i.e. $\Delta B_K^{(2)'}$. 
This gives the estimates that are included in
the error budgets presented in
Refs.~\cite{ref:wlee:2009-2,ref:wlee:2009-3}.

There are various ways in which one can firm up and reduce the
truncation error.
One is to work on a yet finer lattice, which might allow one
to fit to a combination of $a^2$ and $\alpha^2$ errors, and will,
in any case, reduce the size of the error.
Another is to use two-loop matching. And, finally, one can
remove all truncation errors, and replace them with statistical
and some new systematic errors, by using non-perturbative renormalization.
We are pursuing all three approaches.

\begin{table}[htbp]
\centering
\begin{tabular}{ c | c c c c c c}
\hline \hline
$a$ (fm) & $B^{(0)}_K$ & $B^{(1)}_K$ &
$\Delta B_K^{(1)}$ & $\alpha_s(1/a)$ & $\Delta B_K^{(2)}$ & $\Delta B_K^{(2)'}$ \\
\hline
0.12 & 0.6898(59) & 0.5704(58) & 0.1194(83)  & 0.3285 & 0.039 & 0.074\\
0.09 & 0.6118(95) & 0.5256(92) & 0.0862(132) & 0.2729 & 0.024 & 0.046\\
0.06 & 0.5963(83) & 0.5158(80) & 0.0805(115) & 0.2337 & 0.019 & 0.033\\
\hline \hline
\end{tabular}
\caption{Results for tree-level and one-loop parallel-matched $B_K$
using the SU(3) SChPT analysis (fit N-BT7). For all lattice spacings
we use the MILC lattices for which $m_l/m_s=1/5$. 
Also given are the one-loop shift, the value of $\alpha_s$, and
estimates of the two-loop shift.
  \label{tab:dbk2-su3}}
\end{table}
\begin{table}[htbp]
\centering
\begin{tabular}{ c | c c c c c c}
\hline \hline
$a$ (fm) & $B^{(0)}_K$ & $B^{(1)}_K$ &
$\Delta B_K^{(1)}$ & $\alpha_s$ & $\Delta B_K^{(2)}$ & $\Delta B_K^{(2)'}$ \\
\hline
0.12 & 0.6879(53)  & 0.5751(49)  & 0.1128(72)  & 0.3285 & 0.037 & 0.074\\
0.09 & 0.6383(130) & 0.5358(122) & 0.1025(178) & 0.2729 & 0.028 & 0.048\\
0.06 & 0.5829(128) & 0.4937(119) & 0.0892(175) & 0.2337 & 0.021 & 0.032\\
\hline \hline
\end{tabular}
\caption{As for Table~\protect\ref{tab:dbk2-su3} except using
the SU(2) SChPT analysis (4X3Y-NNLO fit).
  \label{tab:dbk2-su2}}
\end{table}

\section{Acknowledgments}
C.~Jung is supported by the US DOE under contract DE-AC02-98CH10886.
The research of W.~Lee is supported by the Creative Research
Initiatives Program (3348-20090015) of the NRF grant funded by the
Korean government (MEST). 
The work of S.~Sharpe is supported in part by the US DOE grant
no.~DE-FG02-96ER40956. Computations were carried out
in part on facilities of the USQCD Collaboration,
which are funded by the Office of Science of the
U.S. Department of Energy.

\end{document}